# Detecting the imprint of a kilonova or supernova in short GRB afterglows

N. Guessoum[1], H. Zitouni[2] & R. Mochkovitch[3]

[1] American University of Sharjah, Physics Department, PO Box 26666, Sharjah, UAE

[2] Laboratoire PTEAM , Faculté des sciences, Université Dr Yahia Fares, Pôle urbain, Médéa, Algeria

[3] Sorbonne Université, CNRS, UMR 7095, Institut dAstrophysique de Paris, 98 bis bd Arago, 75014 Paris, France

Preprint online version: January 29, 2019

**ABSTRACT**

*Context.* Short gamma-ray bursts result from mergers of two neutron stars or from collapsars, but probably at a smaller rate. In the first case, a kilonova occurs while in the second case a Type Ic supernova is expected.

*Aims.* Even if future observations of kilonovae in association with gravitational wave events provide better data, detecting a kilonova during an afterglow follow-up would remain useful for exploring the diversity of the kilonova phenomenon. As supernovae produce a weaker gravitational signal, afterglow follow-up will be the only possible method to find one. In this work, we identify the conditions of the burst energy, external density, kilonova mass, supernova luminosity, that are necessary for the detection of a kilonova or supernova in the follow-up of short GRB afterglows.

*Methods.* We have used a simple kilonova model to obtain the peak luminosities and times as a function of mass, expansion velocity and ejected matter opacity. Afterglow light curves are computed for a uniform medium and a stellar wind, in the kilonova and supernova cases, respectively.

*Results.* We represent, using diagrams of the burst kinetic energy vs. density of the external medium, the domains where the kilonova or supernova at maximum is brighter than the afterglow. In the kilonova case we vary the mass, the jet opening angle and the microphysics parameters; for supernovae, we consider SN 98bw-like and ten times dimmer events, and again vary the jet opening angle and the microphysics parameters.

## 1. Introduction

Short gamma-ray bursts (SGRBs), defined as bursts that last less than two seconds, were clearly identified as a class in the 1990s when the analysis (Kouveliotou et al. 1993; McBreen et al. 1994) of the bursts detected by the BATSE (Burst and Transient Source Experiment) instrument on-board the Compton Gamma-Ray Observatory (CGRO) revealed a clear division of the bursts in terms of duration. Kouveliotou et al. (1993) also showed that the hardness ratio (higher energy to lower energy fluxes) of the bursts anti-correlated with the duration (SGRBs having harder spectra than long ones).

Most importantly, there is now a broad agreement on the general physical processes that produce LGRBs and SGRBs: long bursts result from the collapse of massive stars, followed by a hypernova explosion with jets appearing and carrying relativistic ejecta (the so-called Collapsar Model); short bursts are believed to result mainly from the merger of two compact objects (two neutron stars or one neutron star and a black hole), again leading to the production of two opposite jets emitted by the central engine.

However, the distinction between 'mergers' and short-duration bursts has become important lately as the two-second dividing line between SGRBs and LGRBs has been revisited (Horváth 1998; Mukherjee et al. 1998; Hakkila et al. 2000; Balastegui et al. 2001; Horváth 2002; Chattopadhyay et al. 2007; Horváth 2009; Zhang et al. 2012; Bromberg et al. 2013; Zitouni et al. 2015). It now appears that the classification of a burst as 'short' may depend on the instrument that detects it, as well as on its precise and intrinsic (rest-frame) duration (which then also depends on the burst's redshift). Indeed, some bursts that were classified as short (with a duration of, say, 1.5 seconds) and hence regarded as mergers, might actually be collapsars. A number of papers have thus suggested the existence of an 'intermediate' class of bursts (see e.g. Zitouni et al. (2015)), with various reasons proposed, either physical or instrumental.

Until recently, SGRBs were not the subject of nearly as much interest as LGRBs, mainly for statistical reasons: SGRBs make up only about 10-20% of all bursts, depending on the instrument. The situation has drastically changed now that we have entered the gravitational wave and multi-messenger era, with SGRBs being the confirmed electromagnetic counterparts of NS+NS mergers. The spectacular discovery of GRB 170817A in association with the gravitational signal from the coalescence of two neutron stars has confirmed the general scenario where a short gamma-ray burst is produced with an accompanying kilonova powered by radioactivity from a small amount of material ($10^{-2} - 10^{-1}$ M$_\odot$) ejected at high velocity, $0.1 - 0.3c$.

New detections can be expected when LIGO/Virgo resumes operation in 2019, which will open up the field to the exploration of the diversity of kilonovae in terms of mass, composition, velocity distribution, etc. In parallel to this exciting event, the follow-up of SGRB afterglows at larger distances represents another way to detect kilonovae that would then show up as 'bumps' in the lightcurves. A few convincing candidates have already been found (Tanvir et al. 2013; Yang et al. 2015; Jin et al. 2015), and more can be expected in the future. While the data that can be obtained from these distant kilovae is obviously much less detailed and complete, it can still provide a useful complement to the nearby events in the study of the kilonova phenomenon.

Afterglow follow-up could also reveal an underlying supernova if some SGRBs indeed result from collapsars and not from mergers (Virgili et al. 2011; Bromberg et al. 2013). Finding supernova bumps in SGRB afterglows would confirm this possibility and help to determine whether SGRBs from collapsars exhibit any specific behavior that could be recognized very early on, that is during the prompt or early afterglow phases.

A basic requirement for the detection of a kilonova or a supernova in the follow-up of a SGRB afterglow is that it should have a brightness at least comparable to that of the afterglow at the time of its peak luminosity. We apply this condition to obtain 'visibility diagrams' as a function of the burst energy and density of the external medium for different values of the jet opening angle and microphysics parameters. We consider two kilonova masses, $10^{-2}$ and $10^{-1}$ M$_\odot$ and two supernova energies, SN 98bw-like and 10 times dimmer events. Our results are presented in Sects. 2 and 3, for the kilonova and supernova cases respectively. We discuss the results and present our conclusions in Sect. 4.

## 2. Kilonova detection

### 2.1. Kilonova models: Peak time and peak flux

The kilonova emission results from the diffusion of the energy released by radioactivity throughout the material that is ejected during the coalescence. As the envelope becomes more transparent due to the expansion while the radioactive power decreases, the luminosity initially rises to a maximum before declining.

A variety of kilonova models are available from simple (semi-analytical), spherical, radioactive heating parametrized by a power-law and black-body emission (Li & Paczyński 1998; Kawaguchi et al. 2016; Metzger et al. 2015; Metzger 2017), to very complex ones, that are non spherical, with detailed modelling of radioactive heating and radiative transport (Metzger & Fernández 2014; Wollaeger et al. 2018).

The peak time $t_{v,p}$ and peak flux $F_{v,p}$ in a given spectral band depend both on intrinsic (ejected mass, expansion velocity, composition) and extrinsic (viewing angle, distance, reddening) parameters. Considering a reference model with $m_{\rm ej} = 10^{-2}$ M$_\odot$, $v_{\rm exp} = 0.1\,c$, a uniform composition with a material opacity $\kappa = 10$ cm$^2$.g$^{-1}$ representative of lanthanides, we can use scaling laws to estimate the peak time and peak absolute magnitude for different values of $m_{\rm ej}$, $v_{\rm exp}$ or $\kappa$. We have (Rosswog 2015; Wollaeger et al. 2018)

$$t_{v,\,p} = t_{v,\,p}^{\rm ref} \left( \frac{m_{\rm ej}}{10^{-2}\,M_\odot} \right)^{a_t} \left( \frac{v_{\rm exp}}{0.1\,c} \right)^{b_t} \left( \frac{\kappa}{10\text{ cm}^2.\text{g}^{-1}} \right)^{c_t} \tag{1}$$

and

$$\begin{aligned} M_{v,\,p} = {} & M_{v,\,p}^{\rm ref} + a_M \, Log \left( \frac{m_{\rm ej}}{10^{-2}\,M_\odot} \right) + b_M \, Log \left( \frac{v_{\rm exp}}{0.1\,c} \right) \\ & + c_M \, Log \left( \frac{\kappa}{10\text{ cm}^2.\text{g}^{-1}} \right), \end{aligned} \tag{2}$$

where $t_{v,p}^{\rm ref}$ and $M_{v,p}^{\rm ref}$ are the peak time and peak absolute magnitude for reference values of the parameters.

For a kilonova located at a redshift $z = 0.3$ (the typical redshift we adopt in the results presented below) we can use the peak time and absolute magnitude in the H band ($1.65\,\mu$) to obtain the observed peak time and apparent magnitude in the K band ($2.2\,\mu$). The parameters in Eqs.(1) and (2) are taken from (Wollaeger et al. 2018) and listed in Table 1. Depending on the models, the exponents vary somewhat, but the values we have adopted here are representative.

We can now convert the magnitude and peak time to the K band, adopting $z = 0.3$, which simply gives $t_{K,p} = (1 + z)\,t_{H,p} = 1.3\,t_{H,p}$ and

$$\begin{aligned} m_{K,p}^{\rm ref} = {} & M_{H,p}^{\rm ref} + (5\,LogD - 5) - 2.5\,Log(1 + z) \\ = {} & M_{H,p}^{\rm ref} + 40.7 = 26.7 \end{aligned} \tag{3}$$

with $D = 1566$ Mpc for $z = 0.3$.

To obtain the light curves, we have used a simple semi-analytical model, which assumes spherical symmetry, parametrized radioactive heating and black-body emission. The results are calibrated with the above values of $t_{K,p}$ and $m_{K,p}^{\rm ref}$. They are shown in Fig.1 for the following three cases: two red kilonovae with $\kappa = 10$ cm$^2$.g$^{-1}$, $v_{\rm exp} = 0.2c$ and $M_{\rm KN} = 10^{-2}$, and $10^{-1}$ M$_\odot$; and a composite, two-component model to represent the kilonova which was observed in association with the GW 170817 gravitational wave event. The two components are respectively obtained with $\kappa = 10$ cm$^2$.g$^{-1}$, $v_{\rm exp} = 0.15c$ and $M_{\rm KN} = 0.04$ M$_\odot$ (red kilonova) and $\kappa = 0.5$ cm$^2$.g$^{-1}$, $v_{exp} = 0.3c$ and $M_{KN} = 0.012$ $M_\odot$ (blue kilonova). It can be seen that the model fits the kilonova data (Villar et al. 2017) transported to $z = 0.3$ in the K band from the observed H band light curve, following the method described in Zeh et al. (2004).

|  | $a_{t,M}$ | $b_{t,M}$ | $c_{t,M}$ | ref. value |
|---|---|---|---|---|
| peak time | 0.3 | -0.6 | 0.35 | 2.5 days |
| peak abs. mag. | -0.95 | -1.55 | -1.3 | -14 |

**Table 1.** Exponents and reference values adopted in expressions (1) and (2) for the peak time and absolute magnitude of a kilonova in the H band.

## 2.2. Afterglow light curves: uniform external medium

The expected environment for a short burst resulting from a NS+NS or NS+BH merger is a uniform medium of low density. Due to the long delay between its formation and the coalescence, the system, which is likely to receive a strong kick from the second supernova explosion (Wijers et al. 1992; Kalogera et al. 1996), will migrate towards the galactic halo. Indeed, multi-wavelength fits of short GRB afterglow light curves by Fong et al. (2015) indicate external densities typically ranging from $10^{-5}$ to 1 cm$^{-3}$.

To calculate the afterglow light curves, we essentially followed the prescription of Panaitescu & Kumar (2000), who give analytic expressions for the flux as a function of time under various conditions of burst energy, circumstellar density, and microphysics parameters. To these we added the distance and redshift of the burst and whether the light curve exhibits a jet break. In the optical/near-infrared spectral range, we have $v_i < v < v_c$ where $v_i$, $v$ and $v_c$ are respectively the injection, observing and cooling frequencies. Then, the flux before jet break is given (in mJy) by

$$F_\nu(t) = 10^{(1.36-1.06p)} (1+z)^{(p+3)/4} D_{28}^{-2} \times$$
$$E_{52}^{(p+3)/4} n^{1/2} \epsilon_{e,-1}^{p-1} \epsilon_{B,-3}^{(p+1)/4} \nu_K^{(1-p)/2} t_d^{3(1-p)/4} \qquad (4)$$

which leads to

$$F_\nu(t) = \begin{cases} 0.1 (1+z)^{13/10} D_{28}^{-2} \times \\ E_{52}^{13/10} n^{1/2} \epsilon_{e,-1}^{6/5} \epsilon_{B,-3}^{4/5} \nu_K^{-3/5} t_d^{-9/10} \quad (p = 2.2) \\ 0.05 (1+z)^{11/8} D_{28}^{-2} \times \\ E_{52}^{11/8} n^{1/2} \epsilon_{e,-1}^{3/2} \epsilon_{B,-3}^{7/8} \nu_K^{-3/4} t_d^{-9/8} \quad (p = 2.5) \end{cases} \qquad (5)$$

where $z$ is the burst redshift, $D_{28}$ is the luminosity distance (in units of $10^{28}$ cm), $E_{52}$ is the isotropic kinetic energy (in units of $10^{52}$ erg), $n$ is the density (in cm$^{-3}$) of the external medium; $\epsilon_e$ and $\epsilon_B$ are the microphysics redistribution parameters for the electrons and the magnetic field (in units of $10^{-1}$ and $10^{-3}$ respectively), $p$ is the index of the electron distribution and finally $\nu_K$ and $t_d$ are the frequency (in units of $1.36 \times 10^{14}$ Hz, the K-band frequency) and time (in days) in the observer frame (Panaitescu & Kumar 2000). A jet break is expected at an observed time

$$t_b = 0.46 (1+z) \theta_{-1}^{8/3} \left(\frac{E_{52}}{n}\right)^{1/3} \text{ day} \qquad (6)$$

where $\theta_{-1}$ is the jet opening angle (in units of 0.1 rad). Following the break, the temporal slope equals the spectral index $p$ (Rhoads 1999) and we have

$$F_\nu(t) = 10^{(1.11-1.15p)} (1+z)^{(p+3)/2} D_{28}^{-2} \times$$
$$\theta_{-1}^{2(p+3)/3} E_{52}^{(p+3)/3} n^{(3-p)/12} \epsilon_{e,-1}^{p-1} \epsilon_{B,-3}^{(p+1)/4} \nu_K^{(1-p)/2} t_d^{-p}$$

leading to

$$F_\nu(t) = \begin{cases} 0.038 (1+z)^{13/5} D_{28}^{-2} \times \\ \theta_{-1}^{52/15} E_{52}^{26/15} n^{1/15} \epsilon_{e,-1}^{6/5} \epsilon_{B,-3}^{4/5} \nu_K^{-3/5} t_d^{-2.2} \quad (p = 2.2) \\ 0.015 (1+z)^{11/4} D_{28}^{-2} \times \\ \theta_{-1}^{11/2} E_{52}^{11/6} n^{1/24} \epsilon_{e,-1}^{3/2} \epsilon_{B,-3}^{7/8} \nu_K^{-3/4} t_d^{-2.5} \quad (p = 2.5) \end{cases} \qquad (7)$$

It can be noticed that after the break the flux depends only very weakly on the density of the external medium (as $n^{\frac{3-p}{12}}$).

In addition to the kilonova lightcurves, we represent in Fig.1 the afterglow in the K band for a burst at a redshift $z = 0.3$ with three different values of the isotropic kinetic energy ($10^{50}$, $10^{51}$ and $10^{52}$ erg), four densities of the external medium (from $10^{-3}$ to 1 cm$^{-3}$), two beaming angles (0.1 and 0.2 rad) and fixed microphysics parameters $\epsilon_e = 0.1$ and $\epsilon_B = 10^{-3}$. As the light curves are proportional to the product $\epsilon_B^{(p+1)/4}$ they should be shifted by slightly less than one order of magnitude up or down when $\epsilon_B = 10^{-2}$ or $10^{-4}$.

## 2.3. Visibility diagrams: Kilonova

To allow the detection of a kilonova associated with a short GRB, three main conditions have to be fulfilled: (i) the burst should have a bright enough optical afterglow so that its location can be accurately determined and efficient deep follow-up observations performed[1]; (ii) the burst redshift should be low enough so that the kilonova remains accessible, at least to 4m class (or larger) telescopes; this sets a limit in redshift $z < z_{max} \sim 0.2 - 0.4$, depending on the kilonova luminosity; (iii) the kilonova at its peak should be brighter than the afterglow. About one third of Swift short bursts have an optical afterglow, which represents approximately three events per year. Then, from the distribution in redshift of short bursts (20% of short Swift bursts have a redshift) it can be estimated that again about one third, that is, one event per year is close enough for a potential kilonova detection.

To illustrate these conditions in Fig. 2 we have represented visibility diagrams in $(E, n)$ coordinates. They show the regions where the kilonova at maximum is brighter than the afterglow. We still take $z = 0.3$ but the result does not depend very much on redshift as long as it remains moderate.

We consider two values of the jet opening angle (0.1 and 0.2 rad) and two red kilonovae, with masses of 0.01 and 0.1 $M_\odot$. The blue component in a 'mixed kilonova' as the one following GW170817/GRB170817A might be as bright or even brighter than the red component, but it peaks earlier when the afterglow is brighter and lasts less, so that the kilonova is more likely to be detected when it is red. The full lines are obtained with $\epsilon_e = 0.1$ and $\epsilon_B = 0.01$, while the dashed and dotted lines respectively correspond to the product $\epsilon_e^{6/5} \epsilon_B^{4/5}$ being divided by 10 and 100 (we adopt a value $p = 2.2$ for the index of the electron distribution but assuming a larger $p = 2.5$ yields similar results, the region where the kilonova is visible being slightly more extended since the decline of the afterglow is steeper). Along the diagonal black line the kilonova peaks at the time of the afterglow jet break. Above this line the relative brightness of the kilonova to the afterglow depends very weakly on density ($\propto n^{1/15}$). The size of the zone in the diagrams where a kilonova can be detected appears to increase with its mass (as could be expected), but in case some positive correlation exists between $M_{KN}$ and the burst energy, the fraction of kilonovae that are accessible may possibly depend on mass less strongly.

---

[1] In principle, even without an optical afterglow, a kilonova search could still be performed in the XRT error box if it is small enough or within the host galaxy if it has been identified, but in practice this has been successful for GRB 170817A only.

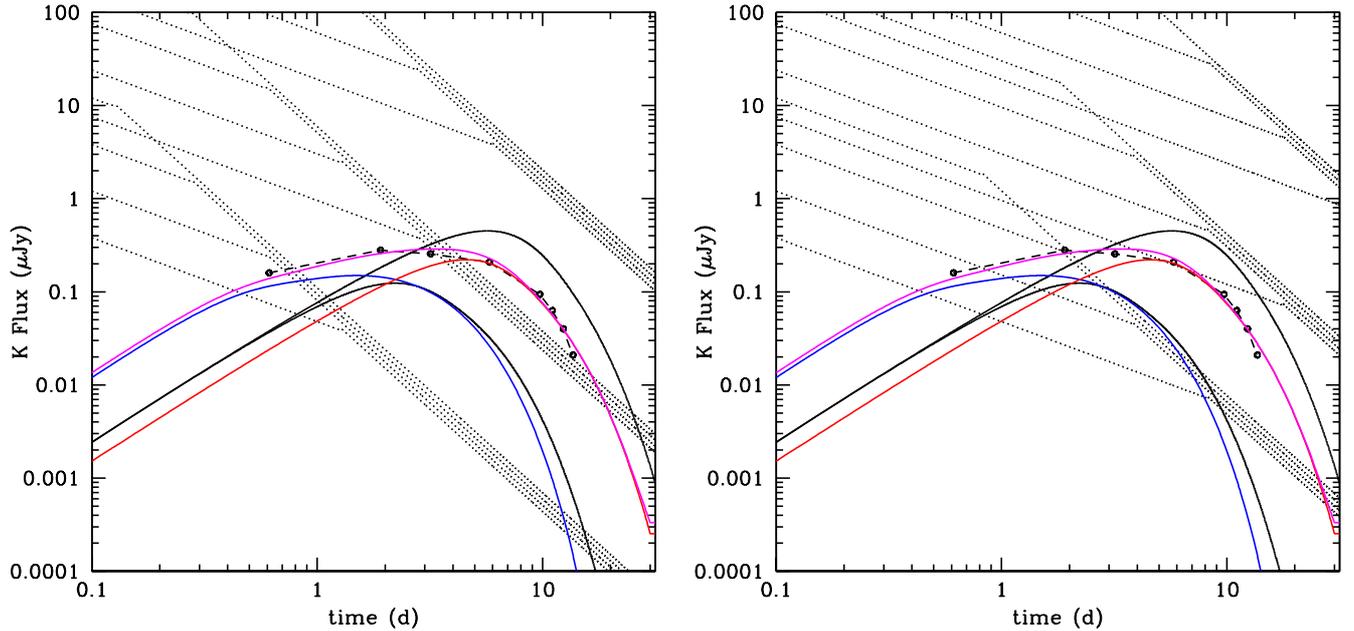

**Fig. 1.** Kilonova and afterglow lightcurves in the K band for two jet opening angles 0.1 rad (left panel) and 0.2 rad (right panel). Three kilonova models are represented: two red ones (full black lines) assuming a uniform opacity $\kappa = 10 \text{ cm}^2.\text{g}^{-1}$, an expansion velocity $v_{exp} = 0.2c$ and masses $M_{KN} = 0.01$ (lower curve) and 0.1 $M_\odot$ (upper curve) and a composite model, sum of a red ($M_{KN} = 0.04 \text{ M}_\odot$, $v_{exp} = 0.15c$, $\kappa = 10 \text{ cm}^2.\text{g}^{-1}$) and a blue ($M_{KN} = 0.012 \text{ M}_\odot$, $v_{exp} = 0.3c$, $\kappa = 0.5 \text{ cm}^2.\text{g}^{-1}$) component (purple, red and blue lines respectively). The dots connected by a dashed line correspond to the light curve of the kilonova associated with GW170817 transported to a redshift $z = 0.3$. The three groups of dotted lines are afterglow light curves for an isotropic kinetic energy of $10^{50}$ (lower), $10^{51}$ (middle) and $10^{52}$ erg (upper group). In each group the four lines correspond to an increasing density of the external medium from $10^{-3}$ to 1 cm$^{-3}$. The adopted values of the microphysics parameters are $\epsilon_e = 0.1$ and $\epsilon_B = 10^{-3}$.

The dots correspond to GRBs in the sample studied by Fong et al. (2015), the burst energy and external density being obtained with $\epsilon_e = 0.1$ and $\epsilon_B = 0.01$ (with the exception of GRB 050724A and GRB 140903A, which were fitted with $\epsilon_B = 10^{-4}$ and $10^{-3}$ respectively). The two green dots represent GRB 050709 and GRB 130603B where an associated kilonova has been found. The cyan (red) dots are bursts located at $z < 0.4$ ($> 0.4$), which we assume to be the horizon for the detection of a bright kilonova. The black dots represent bursts without redshift. Apart from GRB 050709 and GRB 130603B, six events are located at $z < 0.4$ where a kilonova could in principle be accessible: GRB 071227 ($z = 0.38$), GRB 061201 ($z = 0.11$), GRB 080905A ($z = 0.12$), GRB 050724A ($z = 0.258$), GRB 140903A ($z = 0.351$) and GRB 150101B ($z = 0.134$). The GRB 071227 has the lowest isotropic kinetic energy and took place in a high density environment. GRB 061201 and GRB 080905A have similar redshift $\sim 0.1$, energies $E_K \sim 10^{51}$ erg and external densities $n \lesssim 10^{-3} \text{ cm}^{-3}$. GRB 050724A is located close to GRB 130603B in the diagram while GRB 150101B has the lowest external density and a relatively large kinetic energy. Finally, GRB 140903A has the highest isotropic energy and occurred in a low density environment.

GRB 071227 was followed from 0.3 to 40 days in the R band and no kilonova brighter than $R = 24$ was found (D'Avanzo et al. 2009). This is not unexpected since, at the burst distance, a red kilonova with $M_{KN} \leq 0.1 \text{ M}_\odot$ would easily escape detection with a peak magnitude $R > 26.5$. A blue kilonova however (obtained with $\kappa = 0.5 \text{ cm}^2.\text{g}^{-1}$) could have been detected but only if it had a sufficiently large mass, $M_{KN} \sim 0.1 \text{ M}_\odot$, reaching $R \sim 24.3$ at 1.3 days. For GRB 061201 the limits at 1.38 and 3.39 days in the I band ($I > 23.6$ and 24 respectively) are the more constraining (Stratta et al. 2007). They essentially exclude any kilonova (red or blue) more massive than $10^{-2} \text{ M}_\odot$. Similarly for GRB 080905A the limits at 1.5 days in the R band ($R > 25$) exclude a red kilonova more massive than $10^{-2} \text{ M}_\odot$ while a blue kilonova is even more strongly excluded (Rowlinson et al. 2010). In GRB 140903A the upper limits obtained in the $r'$ band at 1.55, 2.5 and 4.49 days (Troja et al. 2016) are typically one order of magnitude above the expected flux of a red, massive kilonova ($M_{KN} = 0.1 \text{ M}_\odot$). A kilonova would have been difficult to detect anyway, even with a deeper search since, due to the large kinetic energy $E_{K,iso} = 4.3^{+1.2}_{-2.0} \times 10^{52}$ erg of this burst, the afterglow is expected to be brighter than the kilonova at its peak as can be seen in Fig. 2. GRB 150101B is discussed in detail by Fong et al. (2016). The most constraining data is the limit $J > 22.3$ at 2.67 days, which imposes an upper mass limit of about 0.1 $M_\odot$ to both a red and blue kilonova. Finally, in the case of GRB 050724A the fluxes measured at 12 and 14h in the K and I band (Berger et al. 2005) were obtained at an early time when the afterglow was bright and the potential kilonova before maximum and therefore too dim to be detected.

More generally the bursts in the Fong et al. (2015) sample fall into two main classes according to the density of the surrounding medium. A first group corresponds to events that occurred in a low density environment ($n \leq 10^{-3}$ cm$^{-3}$). In those bursts the kilonova often peaks before the temporal break in the jet, which makes its detection more difficult. The second group consists of bursts that took place in a denser environment ($n \geq 0.1$ cm$^{-3}$). Then, the jet break occurs earlier making the kilonova potentially more visible. Among those bursts, GRB 071227 and GRB 050724A may have appeared as the best candidates but unfortunately in the first case the search (in the R band) was not deep enough considering that the redshift $z = 0.38$ is relatively large, while in the second case it was performed too early, before the peak of a potential kilonova. In any case, the X-ray light curve followed with Chandra showed no break even after three weeks (Burrows et al. 2009; Grupe et al. 2010), indicating an opening angle larger than 25°, which would have probably made the detection of a kilonova impossible.

## 3. Supernova detection

### 3.1. Supernova models: peak time and peak flux

While it is widely believed that the majority of short GRBs result from the merging of two compact objects, it also appears that a fraction of them can be produced by collapsars (Lazzati et al. 2010; Virgili et al. 2011; Bromberg et al. 2013). In this case, in place of a kilonova one expects that a supernova would show up above the afterglow lightcurve. Detecting a supernova should in principle be easier than a kilonova because supernovae are brighter and peak somewhat later than kilonovae, giving more time for the afterglow to decay. Moreover, the favored environment for a collapsar is a dense stellar wind, which then leads to an early jet break.

As for the kilonova, we used the method of Zeh et al. (2004) to transport supernova light curves in redshift. We considered both bright events, similar to SN 1998bw, and dimmer ones that have a peak luminosity ten times smaller. We adopted a typical redhift $z = 0.3$, but again this value is not critical (as long as it is moderate) in the comparison of the supernova brightness relative to the afterglow. At $z = 0.3$ the observed peak time of the supernova in the $R$ band is 19.8 days with a peak flux of 12.9 $\mu$Jy ($R = 20.9$) for a SN 1998bw-like event.

### 3.2. Afterglow light curves: wind external medium

In the case of a collapsar the most likely nature of the source environnement is the stellar wind of the source progenitor. Then,

before the jet break the afterglow flux in the red band is given (in mJy) by

$$F_\nu(t) = 10^{(2.33-1.23p)} (1+z)^{(p+5)/4} D_{28}^{-2} \times$$
$$E_{52}^{(p+1)/4} A_* \epsilon_{e,-1}^{p-1} \epsilon_{B,-3}^{(p+1)/4} \nu_R^{(1-p)/2} t_d^{(1-3p)/4}, \qquad (8)$$

leading to

$$F_\nu(t) = \begin{cases} 0.42 \, (1+z)^{9/5} \, D_{28}^{-2} \times \\ E_{52}^{4/5} A_* \epsilon_{e,-1}^{6/5} \epsilon_{B,-3}^{4/5} \nu_R^{-3/5} t_d^{-7/5} \; (p=2.2) \\ \\ 0.18 \, (1+z)^{15/8} \, D_{28}^{-2} \times \\ E_{52}^{7/8} A_* \epsilon_{e,-1}^{3/2} \epsilon_{B,-3}^{7/8} \nu_R^{-3/4} t_d^{-13/8} \; (p=2.5) \end{cases} \qquad (9)$$

(Panaitescu & Kumar 2000) where $\nu_R$ is the frequency in units of $4.6 \times 10^{14}$ Hz, the R-band frequency, $A_* = \frac{\dot{M}_{-5}}{4\pi v_8}$ is the wind parameter, $\dot{M}_{-5}$ and $v_8$ being the mass loss rate in units of $10^{-5}$ $M_\odot.$yr$^{-1}$ and the wind velocity in units of $10^8$ cm.s$^{-1}$, respectively. The jet break now takes place at the observer time

$$t_b = 0.034 \, (1+z) \, \theta_{-1}^4 \left( \frac{E_{52}}{A_*} \right) \text{ day} \qquad (10)$$

and the flux after the jet break is

$$F_\nu(t) = 10^{(1.96-1.6p)} (1+z)^{(p+3)/2} D_{28}^{-2} \times$$
$$\theta_{-1}^{p+1} E_{52}^{(p+1)/2} A_*^{(3-p)/4} \epsilon_{e,-1}^{p-1} \epsilon_{B,-3}^{(p+1)/4} \nu_R^{(1-p)/2} t_d^{-p}$$

giving

$$F_\nu(t) = \begin{cases} 0.028 \, (1+z)^{13/5} \, D_{28}^{-2} \times \\ \theta_{-1}^{16/5} E_{52}^{8/5} A_*^{1/5} \epsilon_{e,-1}^{6/5} \epsilon_{B,-3}^{4/5} \nu_R^{-3/5} t_d^{-2.2} \; (p=2.2) \\ \\ 0.009 \, (1+z)^{11/4} \, D_{28}^{-2} \times \\ \theta_{-1}^{7/2} E_{52}^{7/4} A_*^{1/8} \epsilon_{e,-1}^{3/2} \epsilon_{B,-3}^{7/8} \nu_R^{-3/4} t_d^{-2.5} \; (p=2.5) \end{cases} \qquad (11)$$

We can now compare the afterglow flux to the supernova flux at maximum. For

$$A_* > 2.2 \, 10^{-3} \theta_{-1}^4 \, E_{52} \qquad (12)$$

the supernova peaks after jet break, which is satisfied in most cases, except if short GRBs from collapsars can have both a large isotropic energy and a large jet opening angle. This confirms that a supernova, if present, should be easier to detect than a kilonova, being both brighter and peaking at later times, generally after the jet break.

### 3.3. Visibility diagrams: supernova

As for kilonovae, we have represented in Fig.3 visibility diagrams, now in $(E_{52}, A_*)$ coordinates, where the supernova at peak luminosity is brighter than the afterglow. We considered both SN 98bw-like and ten times dimmer events. For both of these we considered three values of the product $\mathcal{E} = \epsilon_{e,-1}^{6/5} \epsilon_{B,-3}^{4/5}$ (10, 1 and 0.1) and three values of the jet opening angle (0.1, 0.2 and 0.3 rad). It can be seen that bright SN 98bw-like events will essentially always be visible for jet opening angles $\theta_j \leq 0.1$. Only for large $\mathcal{E}$ values and large jet opening angles can the afterglow be brighter than the supernova, and only if the burst isotropic kinetic energy is above $10^{52}$ erg. If the supernova is ten times dimmer, the domain where it is visible is somewhat reduced but remains very large, especially for $\theta_j \leq 0.1$, where it still requires an isotropic energy above a few times $10^{52}$ erg for the afterglow to outshine the supernova. The distribution of jet opening angles in short bursts from collapsars remains poorly constrained, but if it is similar to the distribution in long GRBs (Gao & Dai 2010), a majority of events should have $\theta_j \leq 0.2$ rad, making a supernova normally easy to detect at $z \leq 0.3$, under the condition that it is properly searched for, two to three weeks after the burst, when it reaches its peak.

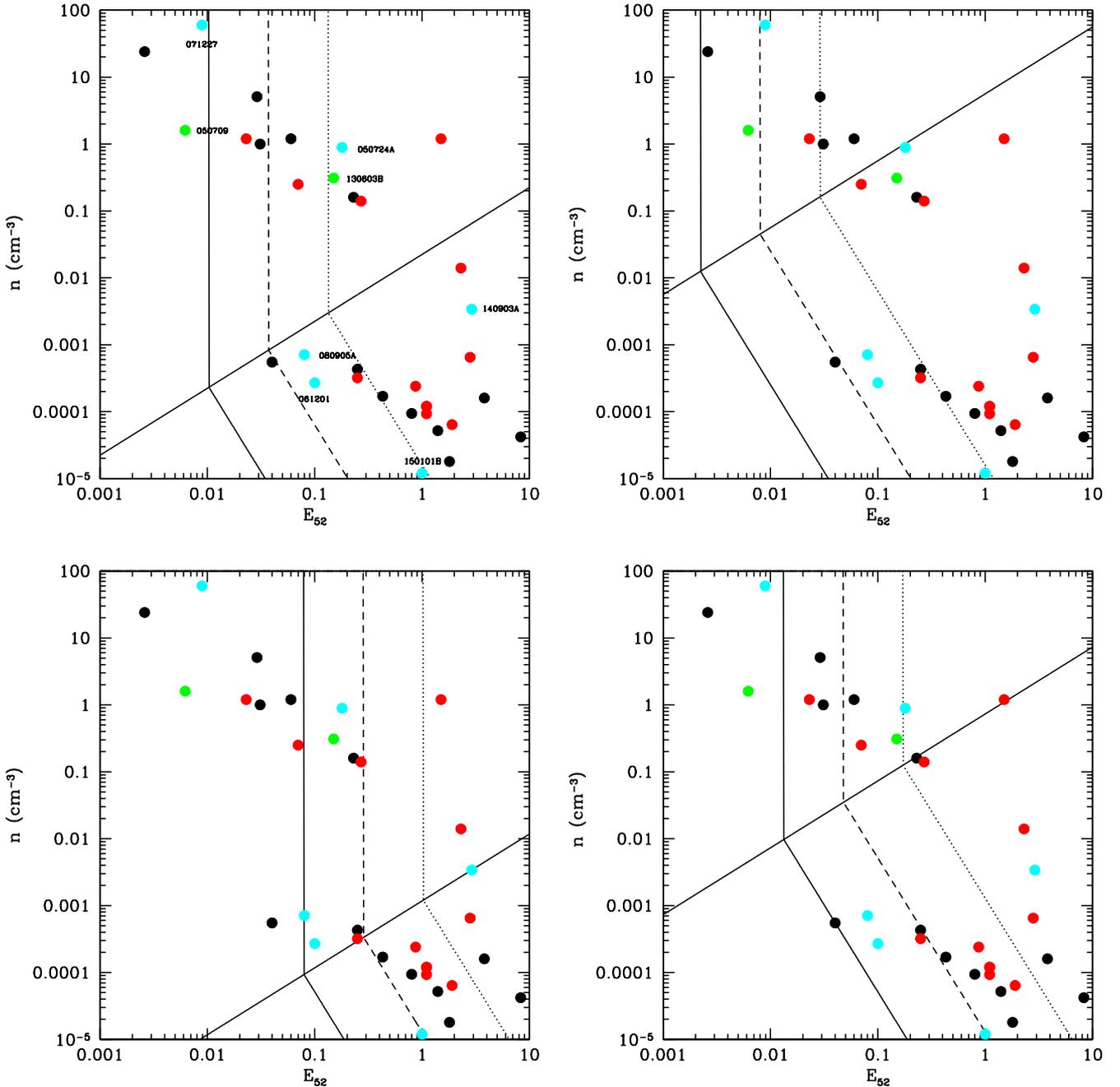

**Fig. 2.** Kinetic energy-external medium density diagrams for short GRBs. The upper (lower) diagrams correspond to $M_{KN} = 10^{-2}$ ($10^{-1}$) M$_\odot$ while the left (right) ones correspond to a jet opening angle $\theta_j = 0.1$ (0.2 rad). In each diagram the diagonal line represents events where the jet break takes place at the peak of the kilonova. The three lines that break when they cross this diagonal, limit the regions where the kilonova is brighter than the afterglow (on the left of the line). The full, dashed and dotted lines correspond to three values of $\mathcal{E} = \epsilon_{e,-1}^{6/5}\epsilon_{B,-3}^{4/5}$: 10, 1 and 0.1. The dots represent GRBs in the sample studied by Fong et al. (2015), in green when a kilonova has been found, in blue (in red) when the measured redshift is below (above) $z = 0.4$, in black when the redshift is not known. In the first diagram, GRBs closer than $z = 0.4$ are identified.

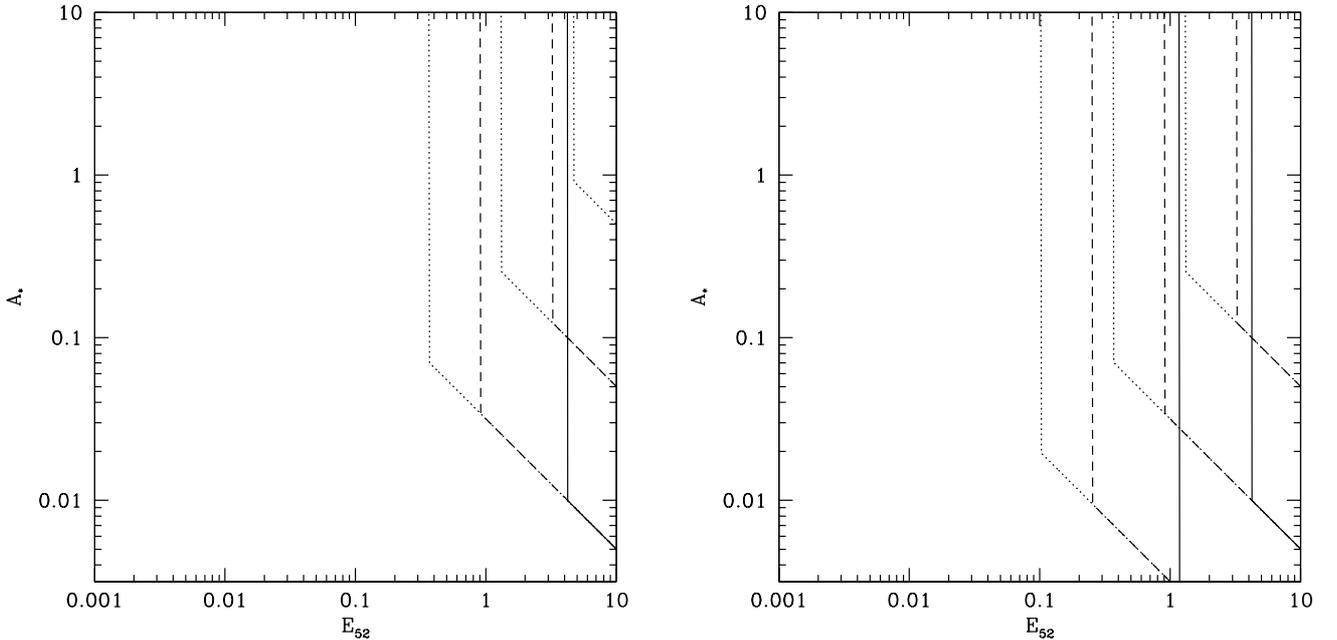

**Fig. 3.** Kinetic energy-wind parameter diagram in the case where the GRB progenitor is a collapsar: left panel SN 98bw-like event; right panel ten times fainter supernova. Three groups of lines are shown corresponding respectively, from bottom to top to $\mathcal{E} = \epsilon_{e,-1}^{6/5} \epsilon_{B,-3}^{4/5} = 10$, 1, and 0.1. In each group the full, dashed and dotted lines are obtained for increasing values of the jet opening angle, $\theta_j = 0.1, 02$, and 0.3 rad. The supernova can be detected above the afterglow in the region of the diagram on the left of each line.

## 4. Discussion and conclusions

Now that we have entered the multi-messenger era, it is probable that most kilonova observations will result from alerts coming from gravitational wave detectors. However, identifying kilonova bumps in afterglow lightcurves, even if it is difficult, will remain interesting and useful to evaluate the diversity in the observational properties of these events. About 20-25% of short bursts with redshift are located at $z < 0.3$ where the detection of a kilonova is in principle possible if it is brighter than the afterglow. The domain where this happens has been represented in the diagrams of Fig. 2. The search for a kilonova should be preferentially made in the infrared (from I to K bands) where it is brighter and stays bright longer. Detecting a blue kilonova component at early times will be even more difficult except if it is much brighter than the one which was associated with GW/GRB 170817. The detection of a kilonova will obviously be easier if the afterglow light curve breaks early, which is favored by a dense environment and/or a small jet opening angle. Indeed, in GRB 130603B the opening angle is constrained within $4 - 8°$ and the kilonova was found after nine days by a deep search at 1.5 μm (F160W HST filter) when at that time the afterglow was below the detection limit. This emphasises the importance of a deep follow-up for more than a week in events which are both sufficiently close ($z \lesssim 0.3$) and exhibit an early jet break. After GRB 130603B two more recent events not included in the Fong et al. (2015) sample, GRB 150424A and GRB 160821B, have been found that satisfy these conditions (Jin et al. 2018): GRB 150424A at $z = 0.3$ with a jet break at 3.8 days does not show any evidence of an underlying kilonova at 1.5 μm down to AB magnitude 26-25 (0.14-0.36 μJy) at five to ten days. This is not very constraining since the afterglow is relatively bright (the inferred isotropic kinetic energy $E_{K,iso} = E_{\gamma,iso}/\epsilon_\gamma$ reaching a few $10^{52}$ erg) and would hide a kilonova less massive than 0.1 $M_\odot$. In GRB 160821B ($z = 0.16$) the situation is less clear but the potential kilonova should have a mass not exceeding $10^{-2}$ $M_\odot$.

Finding a supernova imprint in short GRB afterglows would be very important, as it would confirm that some of the short GRBs are indeed produced by collapsars. Detecting a supernova, if present, should in principle be easy since it peaks later than a kilonova, is brighter and does not require to perform the observations in the infrared. It does, however, require the follow-up of the source to continue for several weeks in bursts closer than $z = 0.3 - 0.5$ (depending on the supernova luminosity), in other words, often after the afterglow has gone below the detection limit. In the very few cases when this follow-up could be performed, no supernova was found and the resulting limits generally exclude any event that would be brighter than one tenth of SN 98bw (Hjorth & Bloom 2012). When more data becomes available, we should be able to constrain the fraction of short GRBs coming from collapsars and to compare the result to the prediction of Bromberg et al. (2013).

With more gravitational wave detections of neutron star mergers and a renewed interest in short GRBs in general we can expect that in the coming years, in addition to several new nearby kilonovae, some more distant ones and possibly some supernovae will be detected in the follow-up of short GRB afterglows. This will allow us to explore whether, in star forming galaxies, short GRBs from mergers and collapsars have specific differences in their prompt and/or early afterglow phases that can be recognized even when no long term follow-up is conducted or is even possible.

## Acknowledgements

NG acknowledges the American University of Sharjah (UAE) for the sabbatical leave and the Institut d'Astrophysique de Paris (France) for hosting him while this work was performed. NG also acknowledges a financial award from the Arab Fund Fellowship Program (Kuwait) and a research grant from the Mohammed Bin Rashid Space Centre (UAE), which supported this work.

## References

Balastegui, A., Ruiz-Lapuente, P., & Canal, R. 2001, MNRAS, 328, 283

Berger, E., Price, P. A., Cenko, S. B., et al. 2005, Nature, 438, 988

Bromberg, O., Nakar, E., Piran, T., & Sari, R. 2013, ApJ, 764, 179

Burrows, D., Garmire, G., Ricker, G., et al. 2009, in Chandra's First Decade of Discovery, ed. S. Wolk, A. Fruscione, & D. Swartz

Chattopadhyay, T., Misra, R., Chattopadhyay, A. K., & Naskar, M. 2007, ApJ, 667, 1017

D'Avanzo, P., Malesani, D., Covino, S., et al. 2009, A&A, 498, 711

Fong, W., Berger, E., Margutti, R., & Zauderer, B. A. 2015, ApJ, 815, 102

Fong, W., Margutti, R., Chornock, R., et al. 2016, ApJ, 833, 151

Gao, Y. & Dai, Z.-G. 2010, Research in Astronomy and Astrophysics, 10, 142

Grupe, D., Burrows, D. N., Wu, X.-F., et al. 2010, ApJ, 711, 1008

Hakkila, J., Haglin, D. J., Pendleton, G. N., et al. 2000, ApJ, 538, 165

Hjorth, J. & Bloom, J. S. 2012, The Gamma Ray Burst Supernova Connection (Cambridge University Press), 169–190

Horváth, I. 1998, ApJ, 508, 757

Horváth, I. 2002, A&A, 392, 791

Horváth, I. 2009, Ap&SS, 323, 83

Jin, Z.-P., Li, X., Cano, Z., et al. 2015, ApJ, 811, L22

Jin, Z.-P., Li, X., Wang, H., et al. 2018, ApJ, 857, 128

Kalogera, V. 1996, ApJ, 471, 352

Kawaguchi, K., Kyutoku, K., Shibata, M., & Tanaka, M. 2016, ApJ, 825, 52

Kouveliotou, C., Meegan, C. A., Fishman, G. J., et al. 1993, ApJ, 413, L101

Lazzati, D., Morsony, B. J., & Begelman, M. C. 2010, ApJ, 717, 239

Li, L.-X. & Paczyński, B. 1998, ApJ, 507, L59

McBreen, B., Hurley, K. J., Long, R., & Metcalfe, L. 1994, MNRAS, 271, 662

Metzger, B. D. 2017, Living Reviews in Relativity, 20, 3

Metzger, B. D., Bauswein, A., Goriely, S., & Kasen, D. 2015, MNRAS, 446, 1115

Metzger, B. D. & Fernández, R. 2014, MNRAS, 441, 3444

Mukherjee, S., Feigelson, E. D., Jogesh Babu, G., et al. 1998, ApJ, 508, 314

Panaitescu, A. & Kumar, P. 2000, ApJ, 543, 66

Rhoads, J. E. 1999, ApJ, 525, 737

Rosswog, S. 2015, International Journal of Modern Physics D, 24, 1530012

Rowlinson, A., Wiersema, K., Levan, A. J., et al. 2010, MNRAS, 408, 383

Stratta, G., D'Avanzo, P., Piranomonte, S., et al. 2007, A&A, 474, 827

Tanvir, N. R., Levan, A. J., Fruchter, A. S., et al. 2013, Nature, 500, 547

Troja, E., Sakamoto, T., Cenko, S. B., et al. 2016, ApJ, 827, 102

Villar, V. A., Guillochon, J., Berger, E., et al. 2017, ApJ, 851, L21

Virgili, F. J., Zhang, B., O'Brien, P., & Troja, E. 2011, ApJ, 727, 109

Wijers, R. A. M. J., van Paradijs, J., & van den Heuvel, E. P. J. 1992, A&A, 261, 145

Wollaeger, R. T., Korobkin, O., Fontes, C. J., et al. 2018, MNRAS, 478, 3298

Yang, B., Jin, Z.-P., Li, X., et al. 2015, Nature Communications, 6, 7323

Zeh, A., Klose, S., & Hartmann, D. H. 2004, ApJ, 609, 952

Zhang, Z. B., Chen, D. Y., & Huang, Y. F. 2012, ApJ, 755, 55

Zitouni, H., Guessoum, N., Azzam, W. J., & Mochkovitch, R. 2015, Ap&SS, 357, 7